\documentclass{IEEE_lsens}
\usepackage{textcomp}
\usepackage{subfigure}
\usepackage[pdftex]{graphicx}
\usepackage[ruled,vlined,linesnumbered]{algorithm2e}
\usepackage{comment}
\usepackage{color}

\usepackage[noadjust]{cite}
\ifCLASSINFOpdf
\else
\fi

\usepackage[T1]{fontenc} %
\usepackage{amsmath}
\usepackage[cmintegrals]{newtxmath}
\usepackage{bm}
\usepackage{array}
\usepackage{url}

\ifCLASSINFOpdf
\else
\fi
\providecommand{\hypersetup}[1]{\relax}

\hypersetup{pdftitle={An Algorithm for Sensor Data Uncertainty Quantification},%
pdfsubject={Typesetting},%
pdfauthor={James T. Meech},%
pdfkeywords={Sensor, Noise, Bayes, Rule, Uncertainty}}%
\begin{document}

\markboth{Vol.~1, No.~3, July~2017}{0000000}

\IEEELSENSarticlesubject{Sensor Signal Processing}

\title{An Algorithm for Sensor Data Uncertainty Quantification}

\author{\IEEEauthorblockN{James~T.~Meech\IEEEauthorrefmark{1}\IEEEauthorieeemembermark{1} and Phillip~Stanley-Marbell\IEEEauthorrefmark{1}\IEEEauthorieeemembermark{2}}%
\IEEEauthorblockA{\IEEEauthorrefmark{1}Department of Engineering, University of Cambridge, Cambridge, CB3 0FA, UK
\IEEEauthorieeemembermark{1}Student Member, IEEE
\IEEEauthorieeemembermark{2}Senior Member, IEEE}%
\thanks{Corresponding authors: James T. Meech and Phillip Stanley-Marbell (e-mail: jtm45@cam.ac.uk, phillip.stanley-marbell@eng.cam.ac.uk).\protect}%
\thanks{Royal society grant RG170136, Alan Turing Institute award TU/B/000096 under EPSRC grant EP/N510129/1, EPSRC grant EP/V047507/1, EPSRC grant EP/V004654/1, and EPSRC grant EP/L015889/1 supported this research.}}
\IEEELSENSmanuscriptreceived{Manuscript received June 7, 2017;
revised June 21, 2017; accepted July 6, 2017.
Date of publication July 12, 2017; date of current version July 12, 2017.}

\IEEEtitleabstractindextext{%
\begin{abstract}
This article presents an algorithm for reducing measurement uncertainty of one physical quantity when given oversampled measurements of two physical quantities with correlated noise.
The algorithm assumes that the aleatoric measurement uncertainty in both physical quantities follows a Gaussian distribution and relies on sampling faster than it is possible for the measurand (the true value of the physical quantity that we are trying to measure) to change (due to the system thermal time constant) to calculate the parameters of the noise distribution.
In contrast to the Kalman and particle filters, which respectively require state update equations and a map of one physical quality, our algorithm requires only the oversampled sensor measurements.
When applied to temperature-compensated humidity sensors, it provides reduced uncertainty in humidity estimates from correlated temperature and humidity measurements. 
In an experimental evaluation, the algorithm achieves average uncertainty reduction of 10.3\,\%.
The algorithm incurs an execution time overhead of 5.3\,\% when compared to the minimum algorithm required to measure and calculate the uncertainty. 
Detailed instruction-level emulation of a C-language implementation compiled to the RISC-V architecture shows that the uncertainty reduction program required 0.05\,\% more instructions per iteration than the minimum operations required to calculate the uncertainty.
\end{abstract}

}

\maketitle

\section{Introduction}
\label{section:introduction}
There are two main types of uncertainty in sensor measurements, aleatoric uncertainty which refers to the uncertainty due to noisy measurements and epistemic uncertainty which refers to the absence of measurements.
We will refer to the noise and offset free true value of the physical quantity that we are trying to measure as the \textit{measurand}.
We will refer to any phenomenon such as temperature, pressure, humidity, and so on, measured using a sensor, as a \textit{physical quantity}.
Many noise reduction techniques rely on filtering or averaging the random variation in sensor signals to reduce the aleatoric uncertainty~\cite{Zhou2020}. 
The lock-in, Kalman, and particle filters are notable exceptions. 
The lock-in filter attenuates all frequencies outside a narrow band around the signal of interest
and must know the expected frequency and phase of the signal~\cite{scofield1994frequency}.  
The Kalman and particle filters combine measurements and estimates of the system state to reduce uncertainty~\cite{kalman1960linear}. 
The Kalman filter calculates the proportion of the weight given to the measurement and the estimate using the covariances of the signals. 
The particle filter calculates the likelihood of individual sensor samples using an estimate of the state~\cite{arulampalam2002tutorial}. 
Our method is fundamentally different to the Kalman and particle filters as it relies on correlation in the noise rather than the signals (as it requires sampling faster than it is physically possible for the measurement to change) and doesn't require the state equations like the Kalman filter or the map of one physical quantity like the particle filter. 
Our approach differs from related work that measures correlation in sensor signals to reduce uncertainty~\cite{gupta2020collaborative,silvestri2018framework} as we apply constraints to ensure that we are only measuring correlation in the noise.  

\begin{figure}[]
	\centering
	\includegraphics[width=0.4\textwidth]{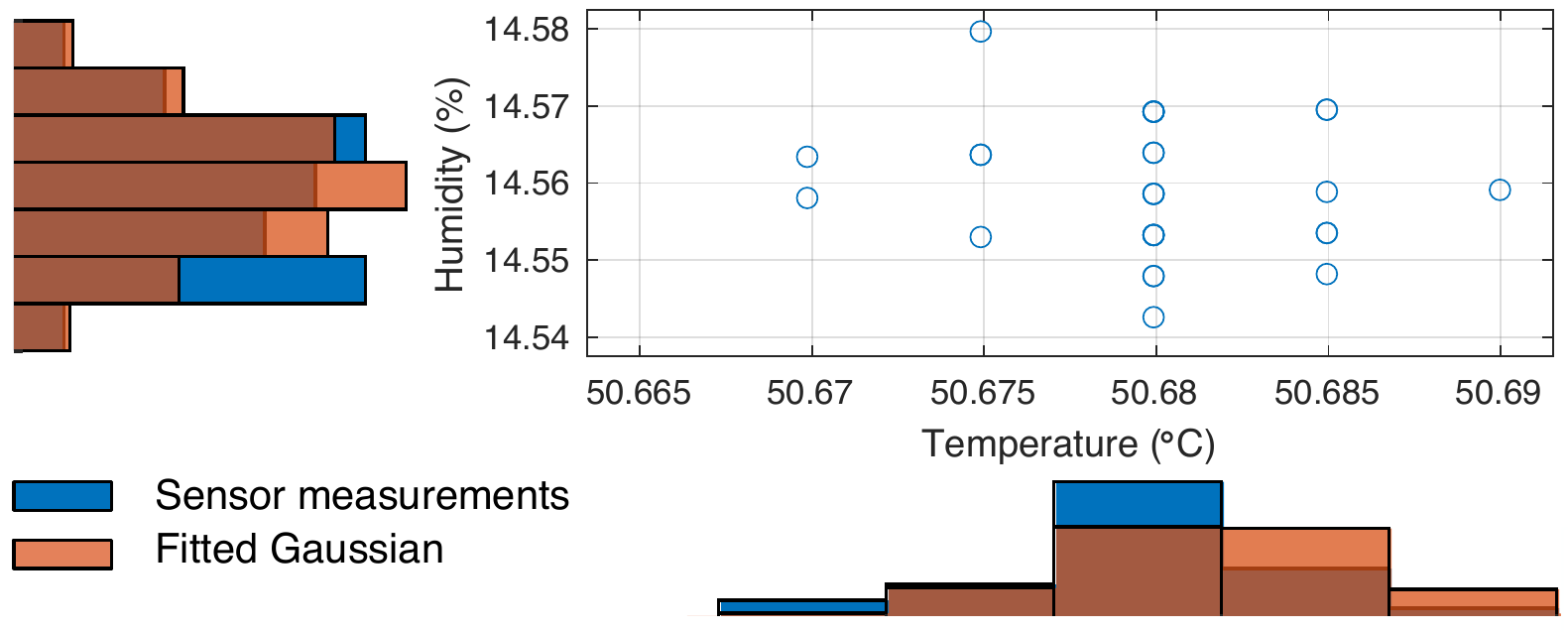}
	\caption{A 30-sample bivariate Gaussian distribution for temperature and humidity measured using a
			 Bosch BME680 sensor at a sample rate of 128.3\,Hz and ambient temperature of 50\,$^\circ$C.
			 Overlaid are histograms of $10^6$ values sampled from their respective fitted Gaussians.}
	\label{figure:principal}
\end{figure}

\noindent \subsubsection{Contribution} This article introduces a new method for reducing the uncertainty of sensor measurements.
The method uses the joint distribution of the noise in two signals to reduce the uncertainty in one of the signals. 
Our method quantifies noise by sampling faster than it is possible for the time-varying measurement signal to change due to physical constraints (the thermal time constant) in the sensor.
Our method assumes that the measurands are constant and uses Bayes rule to calculate a new conditional probability distribution for the most uncertain of the two physical quantities. 
The method checks that each noise distribution is Gaussian with thresholds on skewness and kurtosis. 

\section{Theory and Assumptions}
\label{section:theoryAndAssumptions}
\noindent 
We can model the additive noise in a given temperature measurement as a random variable $T_\mathrm{n}$ with a mean of
$\mu_{T_\mathrm{n}}$ and standard deviation of $\sigma_{T_\mathrm{n}}$. 
Similarly we can model the additive noise in a humidity measurement as a random variable $H_\mathrm{n}$ with a mean of $\mu_{H_\mathrm{n}}$ and a standard deviation of $\sigma_{H_\mathrm{n}}$.
Algorithm~\ref{algorithm:stateOfTheArtUncertainty} shows the minimum operations required to measure and calculate the uncertainty in two coherently-sampled sets of temperature and humidity measurements. 
Figure~\ref{figure:principal} shows a distribution of real sensor data collected using a state-of-the-art 
custom embedded system~\cite{marbell2020warp} and the histograms of samples from their fitted Gaussians. 
We used the BME680~\cite{sensortec2019bme680}, a sensor which measures temperature, pressure, humidity, and indoor air quality to collect the data for Figure~\ref{figure:principal}.
Section~\ref{section:ApplicabilityToOtherSensors} shows that 99.2\,\% of the measurements we took using the BME680 satisfy our empirical thresholds for the data to be considered Gaussian distributed.
The assumption that $T_\mathrm{n}$ (or the noise in any other physical quantity) is Gaussian distributed is only valid if $\sigma_{T_\mathrm{n}} >> \Delta T$ where $\Delta T$ is the change in the measurement
during the time taken to sample the noise distribution~\cite{meech2020efficient}.  
The measurement will follow the value of the measurand with a time lag 
determined by the time constant of the sensor, $\tau_T$. 
In practice, $\tau_T$ places an upper limit on the rate of change of the measurement $\frac{\Delta X}{\Delta t}$. 
Providing that the reciprocal of the sample
rate, $\frac{1}{f_\mathrm{s}} << \tau_T$, the measurement cannot change 
enough to produce a skew in the noise distribution. 
For the BME680, the humidity sensor time constant is $\approx 8$\,s according to the datasheet~\cite{sensortec2019bme680}.
Section~\ref{Section:timeConstantMeasurement} describes an experiment where we measured the time constant of the BME680 temperature sensor on an embedded system~\cite{marbell2020warp} to be $\approx 478$\,s.
All measurements in Section~\ref{section:methodology} satisfy the constraint $\frac{1}{f_\mathrm{s}} << \tau_T$. 
The constraint is an extension of the Nyquist-Shannon sampling theorem.
The BME680's datasheet does not specify the time constant for the pressure sensor and it is difficult to subject the sensor to a step change in pressure to measure it. 
Therefore we do not perform any experiments involving the pressure sensor.

Some modern sensors (such as the BME680 which we use for all the experiments in this work) measure temperature and use it to compensate measurements of other physical quantities such as pressure or humidity~\cite{sensortec2019bme680}. Equations~\ref{equation:temperature}--\ref{equation:pressure} in Section~\ref{section:SensorMeasurementTemperatureCompensation} directly use the temperature measurements as inputs for Equations~\ref{equation:humidityBME680}~and~\ref{equation:pressure} which calculate the humidity and pressure measurements from the BME680 raw ADC values.
As the humidity and temperature sensors are in the same device, there is correlation $\rho_{H_\mathrm{n}T_\mathrm{n}}$ between the random variables $H_\mathrm{n}$ and $T_\mathrm{n}$
due to common supply voltage noise and the use of the temperature measurement
to compensate the humidity calculation in Equation~\ref{equation:humidityBME680}.
Given our assumptions, the joint probability density function of random variables $H_\mathrm{n}$ and $T_\mathrm{n}$ is
\begin{multline}
f_{H_\mathrm{n},T_\mathrm{n}}(h,t) = \frac{1}{2 \pi \sigma_{H_\mathrm{n}} \sigma_{T_\mathrm{n}} \sqrt{1-\rho_{H_\mathrm{n}T_\mathrm{n}}^2}} \mathrm{e}^{-z} \\
z = \frac{\big(\frac{h-\mu_{H_\mathrm{n}}}{\sigma_{H_\mathrm{n}}}\big)^2+\big(\frac{t-\mu_{T_\mathrm{n}}}{\sigma_{T_\mathrm{n}}}\big)^2 - \frac{2 \rho_{H_\mathrm{n}T_\mathrm{n}} (h - \mu_{H_\mathrm{n}}) (t - \mu_{T_\mathrm{n}})}{\sigma_{H_\mathrm{n}} \sigma_{T_\mathrm{n}}}}{2(1-\rho_{H_\mathrm{n}T_\mathrm{n}}^2)}.
\label{equation:normalJointProbability}
\end{multline}

If our assumptions hold and the noise in $H_\mathrm{n}$ and $T_\mathrm{n}$ is Gaussian distributed, our embedded system can:
\begin{enumerate}
	\item Take $N$ pairs of samples from the humidity and temperature signals with $\frac{1}{f_\mathrm{s}} << \tau_T$.
	\item Assume the noise is Gaussian distributed and additive.
	\item Check that the skewness and kurtosis of the noise are within a threshold where this assumption is acceptable. 
	\item Fit the sampled noisy measurements to Equation~\ref{equation:normalJointProbability}.
	\item Use Bayes rule and assert that $t = \mu_{T_\mathrm{n}}$ to obtain a new conditional probability distribution for the humidity:
\end{enumerate}
\begin{multline}
f_{H_\mathrm{n},T_\mathrm{n}}(h|t=\mu_{T_\mathrm{n}}) = \frac{f_{H_\mathrm{n},T_\mathrm{n}}(t=\mu_{T_\mathrm{n}},h)}{f_{H_\mathrm{n},T_\mathrm{n}}(t=\mu_{T_\mathrm{n}})} \\
f_{H_\mathrm{n},T_\mathrm{n}}(h|t=\mu_{T_\mathrm{n}}) = \frac{1}{\sqrt{2\pi}\hat{\sigma}_{H_\mathrm{n}}} \mathrm{e}^{-\frac{1}{2}\big(\frac{h-\mu_{H_\mathrm{n}}}{\hat{\sigma}_{H_\mathrm{n}}}\big)^2} \\
\hat{\sigma}_{H_\mathrm{n}} = \sigma_{H_\mathrm{n}} \sqrt{1 - \rho^2_{H_\mathrm{n}T_\mathrm{n}}}.
\label{equation:integrationConstant}
\end{multline}

To ensure that the assumptions are valid for each distribution the method checks that kurtosis $<7$ and that the $|$skew$|$ $<2$~\cite{west1995structural}. 
These thresholds are widely referred to as a good rule-of-thumb check that a set of samples are Gaussian distributed.
If the noise violates either of these conditions then the method will report the $\sigma_{H_\mathrm{n}}$ instead of the conditional $\hat{\sigma}_{H_\mathrm{n}}$. 
Algorithm~\ref{algorithm:onSensorInference} shows the new method we propose for calculating the humidity uncertainty at a fixed temperature using Equation~\ref{equation:integrationConstant}.

\begin{algorithm}[]
\DontPrintSemicolon
\SetAlgoLined
\KwResult{$\mu_{T_\mathrm{n}}$, $\mu_{H_\mathrm{n}}$, $\sigma_{T_\mathrm{n}}$, $\sigma_{H_\mathrm{n}}$}
Collect $N$ samples from $H_\mathrm{n}$\;
Collect $N$ samples from $T_\mathrm{n}$\; 
$\mu_{H_\mathrm{n}}$ = Mean($H_\mathrm{n}$)\;
$\mu_{T_\mathrm{n}}$ = Mean($T_\mathrm{n}$)\;
$\sigma_{H_\mathrm{n}}$ = Standard deviation($H_\mathrm{n}$)\;
$\sigma_{T_\mathrm{n}}$ = Standard deviation($T_\mathrm{n}$)\;
\caption{The bare minimum set of operations required for uncertainty measurement and quantification.}
\label{algorithm:stateOfTheArtUncertainty}
\end{algorithm}

\begin{algorithm}[]
\DontPrintSemicolon
\SetAlgoLined
\KwResult{$\mu_{T_\mathrm{n}}$, $\mu_{H_\mathrm{n}}$, $\sigma_{T_\mathrm{n}}$, $\hat{\sigma}_{H_\mathrm{n}}$}
Collect $N$ samples from $T_\mathrm{n}$ and $H_\mathrm{n}$ at sample rate $f_\mathrm{s}$\; 
$\mu_{H_\mathrm{n}}$ = Mean($H_\mathrm{n}$)\;
$\mu_{T_\mathrm{n}}$ = Mean($T_\mathrm{n}$)\;
$\sigma_{H_\mathrm{n}}$ = Standard deviation($H_\mathrm{n}$)\;
$\sigma_{T_\mathrm{n}}$ = Standard deviation($T_\mathrm{n}$)\;
\uIf{($|$skewness$|$ $<$ 2) \& (kurtosis $<$ 7)}{
  	$\rho_{H_\mathrm{n}T_\mathrm{n}}$ = Correlation($H_\mathrm{n}$, $T_\mathrm{n}$)\; 
	$\hat{\sigma}_{H_\mathrm{n}} = \sigma_{H_\mathrm{n}} \sqrt{1-\rho_{H_\mathrm{n}T_\mathrm{n}}^2}$\; 	
	}
\Else{
  $\hat{\sigma}_H$ = $\sigma_H$\;
  }
\caption{Our new proposed method for uncertainty measurement and quantification at a given temperature.}
\label{algorithm:onSensorInference}
\end{algorithm}
\subsection{Applicability to Other Sensors}
\label{section:ApplicabilityToOtherSensors}
\begin{figure}[]
	\centering
	\includegraphics[width=0.44\textwidth]{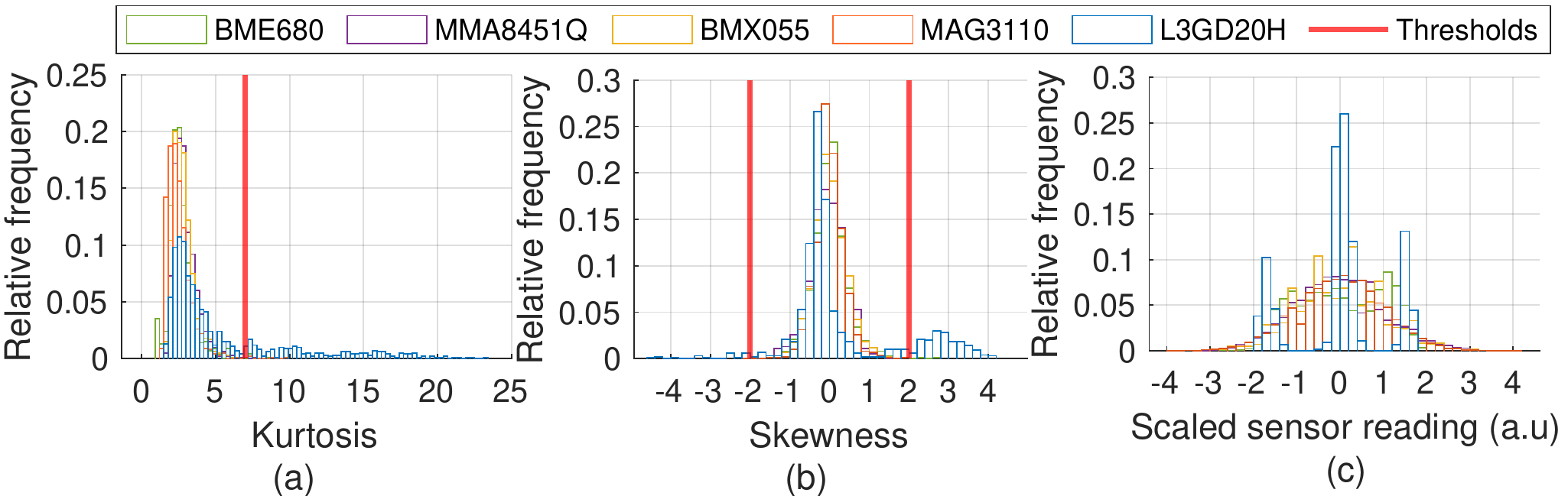}
	\caption{Histograms of the kurtosis (a) and skewness (b) of 333 thirty-sample distributions from all axes / physical quantities measured by each sensor compacted into one histogram per sensor. 
     (c) Histogram of 10,000 sample stationary distributions of the inertial sensor z-axes and the BME680 pressure scaled to fit them in the same domain.}
	\label{figure:empiricalStudy}
\end{figure}

\noindent We conducted an empirical sensor applicability study using an embedded sensor system~\cite{marbell2020warp}.
The sensors include two 3-axis accelerometers (MMA8451Q, BMX055), a gyroscope (L3GD20H), a magnetometer (MAG3110), and an environmental sensor (BME680).
We placed the sensor system on a vibration isolation stage in a thermal chamber set to 25\,$^\circ$C with a sensor supply voltage of 2.5\,V, waited for one hour for the temperature to equilibrate, then sampled 10,000 samples from each sensor at a sample rate of 4.9\,Hz.  
Figures~\ref{figure:empiricalStudy}(a)~and~\ref{figure:empiricalStudy}(b) show the kurtosis and skewness of 333 thirty-sample distributions from each axis or physical quantity from each sensor. 
We calculated that 75.8\,\% of the L3GD20H data, 99.6\,\% of the MAG3110 data, 100\,\% of the BMX055 data, 99.9\,\% of the MMA8451Q data, and 99.2\,\% of the BME680 data are within the 	threshold.
Figure~\ref{figure:empiricalStudy}(c) shows the 10,000 sample stationary noise distributions from the z-axis of each inertial sensor and the BME680 pressure sensor. 
The noise distribution of the L3GD20H is a trimodal Gaussian which causes the distributions to fail to meet the skewness and kurtosis thresholds significantly more than the other sensors.

\subsection{How is This Different to Kalman and Particle Filters?}

\noindent 
\textbf{\textit{The Kalman and particle filters measure correlation in the sensor signals whereas Algorithm~\ref{algorithm:onSensorInference} samples faster than it is physically possible for the measurand to change and measures correlation in the noise.}}
Algorithm~\ref{algorithm:onSensorInference} calculates the standard deviation of the noise at a given temperature and
allows the noise to depend on the measurand and other environmental conditions whereas the Kalman filter does not. 
Both the Kalman and particle filters require equations that describe the relationship between the sensor signals, our method requires only the set of oversampled temperature measurements.
The Kalman filter applies no online checks on the shape of the distribution. 
In contrast, Algorithm~\ref{algorithm:onSensorInference} applies checks on the skewness and kurtosis. 
The particle filter can support multimodal distributions but Algorithm~\ref{algorithm:onSensorInference} cannot support anything other than a unimodal Gaussian distribution. Unlike the particle filter Algorithm~\ref{algorithm:onSensorInference} has no predict step and instead measures correlation in the noise.

\noindent \subsubsection{Computational Complexity}

Let $N_\mathrm{O}$ be number of output states in our system and $M$ be the number of weights. 
Arulampalam et. al. and Haykin state that the computational and memory complexity of both the extended Kalman filter and the particle filter is $\mathcal{O}(N_\mathrm{O}M^2)$~\cite{haykin2004kalman,arulampalam2002tutorial}.
We experimentally measured the computational complexity of Algorithm~\ref{algorithm:onSensorInference} to be $\mathcal{O}(N_\mathrm{O}M)$ by plotting the number of instructions required to run the algorithm for different values of $N$.
This is because we calculate the correlation between each variable and the one we are interested in rather than all correlations between all variables which are required by the Kalman and particle filters.
Assuming that the hardware complexity scales with the computational complexity, Algorithm~\ref{algorithm:onSensorInference} offers considerably lower hardware complexity than the Kalman and particle filters.

\subsubsection{Comparison Against the Kalman and Particle Filters}

We cannot perform a like for like empirical experimental complexity comparison with the Kalman or particle filters for this application.
The Kalman filter requires equations relating the humidity and temperature measurements to some shared state. In this application we do not have such equations. 
We cannot perform a like for like empirical experimental comparison with the particle filter because we have no map of how the humidity will change over time.

\subsection{Sensor Measurement Temperature Compensation}
\label{section:SensorMeasurementTemperatureCompensation}
\noindent The BME680~\cite{sensortec2019bme680} uses an internal temperature sensor to compensate its pressure, humidity, and air quality measurements. 
Let $K_{T1}$--$K_{T3}$, $K_{H1}$--$K_{H7}$, and $K_{P1}$--$K_{P10}$ be calibration constants stored inside the sensor's digital logic 
and $T_\mathrm{ADC}$ be the value read by the analogue to digital converter (ADC) of the temperature sensor in the BME680.
We translated the functions \texttt{calc\_temperature}, \texttt{calc\_humidity}, and \texttt{calc\_pressure} provided by Bosch Sensortech~\cite{bme680API} into equations.
The converted temperature
\begin{equation}
T_\mathrm{Out} = \frac{K_{T2}}{5120} \bigg( \frac{T_\mathrm{ADC}}{2^{14}} - \frac{K_{T1}}{2^{10}} \bigg) + \frac{K_{T3}}{5120 \times 2^{10}}\bigg( \frac{T_\mathrm{ADC}}{2^{14}} - \frac{K_{T1}}{2^{10}} \bigg)^2.
\label{equation:temperature}
\end{equation}

Let $H_\mathrm{C}$ be temperature correction coefficient used to compensate the humidity calculation where
\begin{equation}
H_\mathrm{C} = \frac{K_{H2}}{2^{18}} + \frac{K_{H2} K_{H4}}{2^{32}} T_\mathrm{Out} + \frac{K_{H2}K_{H5}}{2^{38}} T^2_\mathrm{Out}.
\label{equation:tcompBME680}
\end{equation}

Let $H_\mathrm{ADC}$ be the value read by the ADC of the humidity sensor where the temperature-compensated humidity
\begin{multline}
H_\mathrm{Out} = H_\mathrm{C} \bigg( H_\mathrm{ADC} - K_{H1} 2^{4} + \frac{K_{H3}}{2} T_\mathrm{Out} \bigg) + \\ 
\bigg( \frac{K_{H6}}{2^{14}} + \frac{K_{H7} T_\mathrm{Out}}{2^{21}} \bigg) H^2_\mathrm{C} \bigg( H_\mathrm{ADC} - K_1 2^{4} + \frac{K_{H3}}{2} T_\mathrm{Out} \bigg)^2.
\label{equation:humidityBME680}
\end{multline}

Equations~\ref{equation:tcompBME680}~and~\ref{equation:humidityBME680} show that any noise in the temperature measurement will propagate through to the humidity measurement.
Let $P_\mathrm{C1}$ and $P_\mathrm{C2}$ be the temperature correction coefficients used to compensate the pressure calculation where
\begin{multline}
P_\mathrm{C1} = \frac{K_{P6}}{2^{19}} \bigg( 2560T_\mathrm{Out} - 64000\bigg)^2 + \\
K_{P5}(1280T_\mathrm{Out}- 32000) + 2^{16}K_{P4}
\label{equation:pressurecomp1}
\end{multline}
\begin{multline}
P_\mathrm{C2} = \frac{K_{P1}K_{P3}}{2^{48}} (2560T_\mathrm{Out} -64000)^2 + \\
 \frac{K_{P1}K_{P2}}{2^{34}}(2560T_\mathrm{Out} - 64000) + K_{P1} 
\label{equation:pressurecomp2}
\end{multline}

Let $P_\mathrm{C3}$ be an intermediate variable in the pressure calculation where
\begin{equation}
P_\mathrm{C3} = \frac{6250}{P_\mathrm{C2}} \bigg( 2^{20} - P_\mathrm{ADC} - \frac{P_\mathrm{C1}}{2^{12}} \bigg).
\label{equation:pressurecomp3}
\end{equation}

Let $P_\mathrm{ADC}$ be the value read by the ADC of the pressure sensor where
\begin{equation}
    P_\mathrm{Out} =
\begin{cases}
    \frac{K_{P10}P_\mathrm{C3}^3}{2^{45}} + \frac{K_{P9}P_\mathrm{C3}^2}{2^{35}} +\\
	 P_\mathrm{C3}\bigg(1 + \frac{K_{P8}}{2^{19}} \bigg) + 2^3 K_{P7},& \text{if } P_{C2} \neq 0\\
    2^{20} - P_\mathrm{ADC},              & \text{otherwise.}
\label{equation:pressure}
\end{cases}
\end{equation}
\section{Methodology} 
\label{section:methodology}

\noindent\subsubsection{Motivation for Experiments} The output of high accuracy metal-oxide gas sensors depends on relative humidity~\cite{Moseley_2017}.
These sensors need to measure extremely small gas concentrations (10 ppm), decreasing the noise in the humidity measurement will enable more accurate compensation for humidity.
\subsection{Temperature Sensor Experiments}
\label{Section:timeConstantMeasurement}
\noindent 
To calibrate the temperature sensor we placed a sensor system that contains a BME680 in a thermal chamber and set the temperature to 0\,$^\circ$C.
We had the sensor system sample at a constant temperature for one hour before performing the calibration to allow the temperature to reach a steady state. 
We had the embedded system sample 12,000 values and repeated this at each temperature from 0 to 40\,$^\circ$C with a step of 5\,$^\circ$C. 
The thermal time constant is the time required for a system subjected to a step change in temperature to reach $1-e^{-1}$ of 
its final temperature~\cite{Pourmovahed1990Experimental}.
We need to measure the thermal time constant of the BME680 on the sensor system to ensure that we satisfy the sampling rate constraint from Section~\ref{section:theoryAndAssumptions}.
To measure the time constant of the temperature sensor we subjected the sensor system to a step change in temperature from 50\,$^\circ$C to room temperature (Figure~\ref{figure:temperatureCalibration}(a)). We repeated the experiment three times.
\subsection{Uncertainty Quantification Experiments}
\noindent We placed two identical sensor systems in a typical indoor environment running 
Algorithm~\ref{algorithm:onSensorInference} at 22.3\,Hz and Algorithm~\ref{algorithm:stateOfTheArtUncertainty}
at 23.6\,Hz for one hour. 
We had one sensor system calculate the mean and standard deviation of the measurement distribution using Algorithm~\ref{algorithm:stateOfTheArtUncertainty} 
and the other calculate the uncertainty at the mean temperature using Algorithm~\ref{algorithm:onSensorInference}.
{We repeated this experiment five times.
Figure~\ref{figure:experimentalSetup} shows the experimental setup that we used to measure the power consumed by each 
sensor system simultaneously. In a separate experiment we used the Sunflower embedded system emulator to emulate both C-language implementations to obtain 
estimates of the number of instructions required to run them on a RISC-V processor
with a sensor sample rate of 128.3\,Hz and a clock speed of 60\,MHz~\cite{Stanley-Marbell:2001:FFC}. 

\begin{figure}[]
	\centering
	\includegraphics[width=0.45\textwidth]{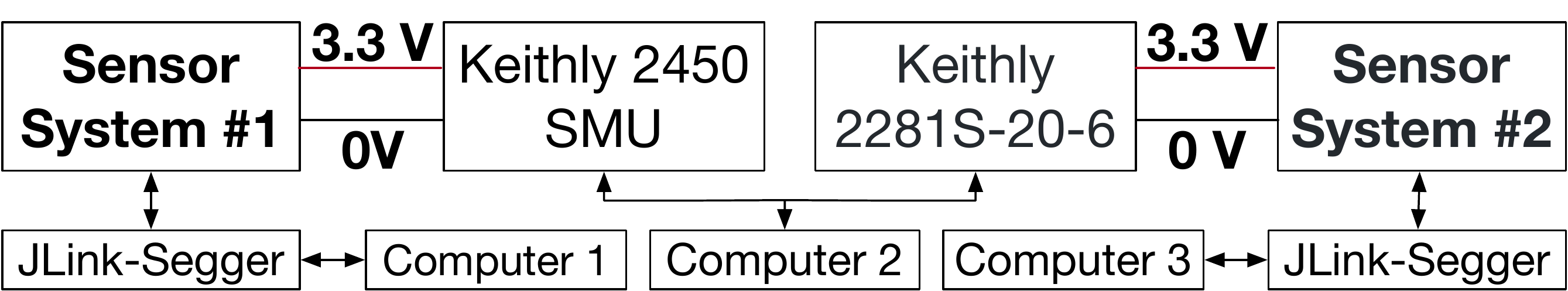}
	\caption{We powered both sensor systems separately so we could measure the power consumption of each algorithm individually.}
	\label{figure:experimentalSetup}
\end{figure}

\section{Results and Discussion}

\subsection{Temperature Sensor Experiments}

\noindent Figure~\ref{figure:temperatureCalibration}(a) shows the response of the sensor to a step change in temperature from 50\,$^\circ$C to 
room temperature. We measured the thermal time constant of the temperature sensor to be 478\,s.
Figure~\ref{figure:temperatureCalibration}(b) shows the result of performing a least-squares regression to map the values from the sensor to the thermal chamber set temperature. 
We assume that the thermal chamber set temperature is the ground truth for the temperature measurand. 
We fit a linear model to remove the temperature offset due to the heating from other components. 
The regression resulted in a best fit line of $T_\mathrm{Measurand} = 1.001 \times T_\mathrm{Sensor} - 2.179$.

\begin{figure}[]
	\centering
	\includegraphics[width=0.45\textwidth]{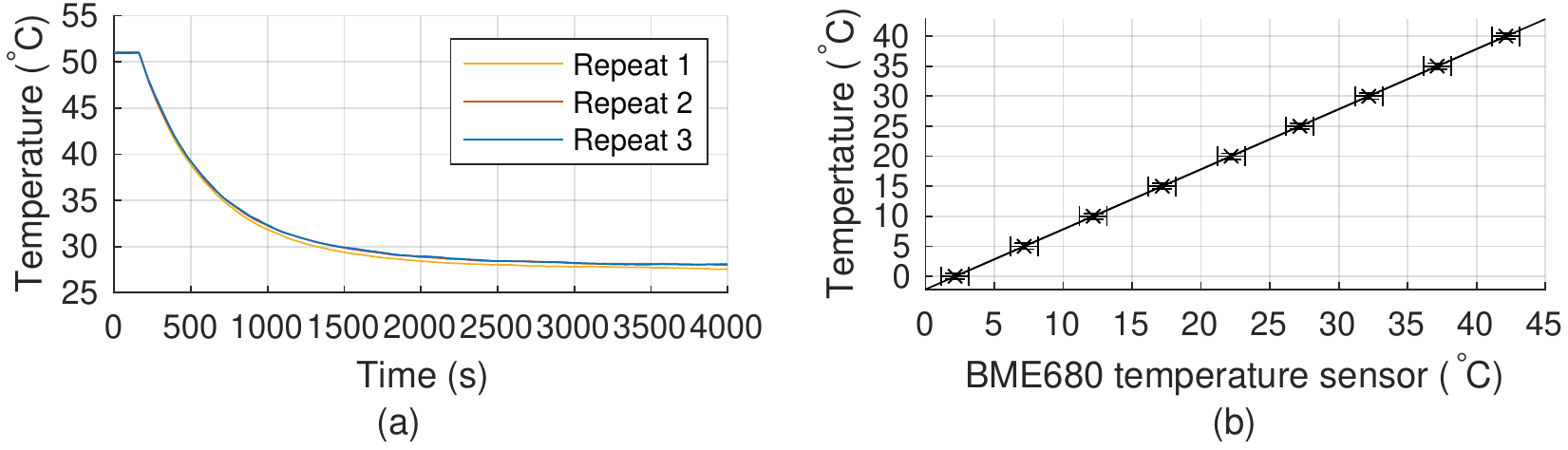}
	\caption{
	(a) Sensor system \#1 time constant measurement.
	(b) Calibration of the BME680 temperature sensor on sensor system \#1 where 
	we calculated the constants for the equation $y = mx + c$ 
    relating the temperature measurement to the temperature measurand by 
	fitting a least-squares line to the data where $c = -2.179$ and $m = 1.001$.
	We plotted the error bars using uncertainties of 0.5 and 1.0\,$^\circ$C
	for the thermal chamber and BME680 respectively~\cite{sensortec2019bme680,thermalCalibration}.}
	\label{figure:temperatureCalibration}
\end{figure}
\subsection{Uncertainty Quantification Experiments}

\noindent Algorithm~\ref{algorithm:onSensorInference} was able to reduce the standard deviation of humidity measurements by 10.3\,\% on average and to do this for 95.4\,\% of the experimental data. 
Figure~\ref{figure:skewnessKurtosis} shows the distribution of the skewness 
and kurtosis for each of the 30 sample distributions for both temperature and humidity for a separate five-hour dataset. 
Figure~\ref{figure:skewnessKurtosis} shows that the data rarely exceeded the thresholds on skewness and kurtosis.
The conditions on line 6 of Algorithm~\ref{algorithm:onSensorInference} are satisfied for 97.9\,\% of the data. 
Algorithm~\ref{algorithm:onSensorInference} requires 2.4\,\% less energy than Algorithm~\ref{algorithm:stateOfTheArtUncertainty} if we run both algorithms for the same period of time. 
Algorithm~\ref{algorithm:stateOfTheArtUncertainty} runs faster, samples from the sensor more in a given amount of time and therefore consumes more power on average. 
The implementation of Algorithm~\ref{algorithm:onSensorInference} required 5.3\,\% more time to run per iteration than that of Algorithm~\ref{algorithm:stateOfTheArtUncertainty}. 

\begin{figure}[]
	\centering
	\includegraphics[width=0.45\textwidth]{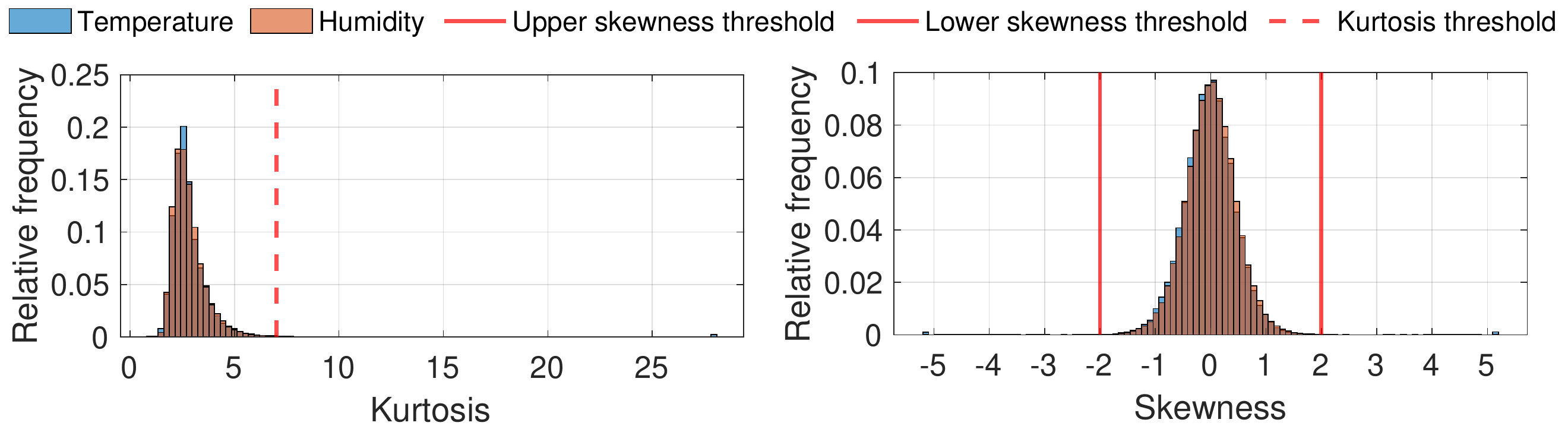}
	\caption{BME680 temperature and humidity skewness and kurtosis for a five hour sample of data in a typical indoor environment.
			 The sensor system calculated skewness and kurtosis for 30 sample distributions.   
			 Skewness and kurtosis values for a perfect Gaussian are zero and three respectively.}
	\label{figure:skewnessKurtosis}
\end{figure}

After running 100 iterations of each C implementation on the processor emulator for 30 sample distributions the instruction-level program emulation showed that the C-implementation of Algorithm~\ref{algorithm:onSensorInference} required 0.05\,\% more instructions to run than that of Algorithm~\ref{algorithm:stateOfTheArtUncertainty}. 
The number of RISC-V instructions executed by the C-language implementations is independent of the system clock speed and therefore generalises to any typical modern architecture.
The instruction costs for both Algorithm~\ref{algorithm:stateOfTheArtUncertainty}~and~\ref{algorithm:onSensorInference} 
scale with $\mathcal{O}(N)$. 
\section{Conclusion}

\noindent This article presents an algorithm that can calculate the aleatoric uncertainty of sensor measurements at a fixed temperature by sampling faster than it is physically possible for the measurement to change to characterise the noise. 
Our method (Algorithm~\ref{algorithm:onSensorInference}) reduced the uncertainty in humidity measurements by 10.3\,\% on average for a speed decrease of 5.3\,\%. 
The speed decrease leads to Algorithm~\ref{algorithm:onSensorInference} requiring 2.4\,\% less energy to run than the baseline (Algorithm~\ref{algorithm:stateOfTheArtUncertainty}) for the same time period. 
Our approach requires less power than the state of the art which will make it attractive to embedded systems engineers working with power-constrained devices.
The performance of Algorithm~\ref{algorithm:onSensorInference} is not limited to an uncertainty reduction of 10.3\,\% for all applications.
In applications where the noise has larger correlation or is sampled at a higher rate we are likely to see larger uncertainty reductions.
Stronger or additional constraints on the value of the measurand of one of the physical quantities would lead to further uncertainty reduction. 
A weakness of the method is the requirement for the sensor noise to be Gaussian. For some sensors such as the L3GD20H this requirement is not satisfied.
Extending the method for the case where the sensor noise is not Gaussian but instead some arbitrary $N$ dimensional distribution where $N$ is the number of sensor signals is an open research problem.

\bibliography{ms}
\bibliographystyle{IEEEtran}

\end{document}